\documentclass[%
reprint,
superscriptaddress,
 amsmath,amssymb,
 aps,
prl,
]{revtex4-1}

\usepackage{graphicx}
\usepackage{dcolumn}
\usepackage{bm}

\usepackage{color}
\usepackage{colordvi}  
\definecolor{Red}{rgb}{1,0,0}

\makeatletter
\def\authornote{\xdef\@thefnmark{$\dagger$}\@footnotetext}
\makeatother

\begin{document}

\title{Self-homodyne enabled generation of indistinguishable photons}

\author{Kai M\"uller$^\dagger$}
\affiliation{E. L. Ginzton Laboratory, Stanford University, Stanford, California 94305, USA}
\author{Kevin A. Fischer$^\dagger$}
\affiliation{E. L. Ginzton Laboratory, Stanford University, Stanford, California 94305, USA}
\author{Constantin Dory$^\dagger$}
\authornote{These authors contributed equally.}
\affiliation{E. L. Ginzton Laboratory, Stanford University, Stanford, California 94305, USA}
\author{Tomas Sarmiento}
\affiliation{E. L. Ginzton Laboratory, Stanford University, Stanford, California 94305, USA}
\author{Konstantinos G. Lagoudakis}
\affiliation{E. L. Ginzton Laboratory, Stanford University, Stanford, California 94305, USA}
\author{Armand Rundquist}
\affiliation{E. L. Ginzton Laboratory, Stanford University, Stanford, California 94305, USA}
\author{Yousif A. Kelaita }
\affiliation{E. L. Ginzton Laboratory, Stanford University, Stanford, California 94305, USA}
\author{Jelena Vu\v{c}kovi\'c}
\email{jela@stanford.edu}
\affiliation{E. L. Ginzton Laboratory, Stanford University, Stanford, California 94305, USA}

\date{\today}

\onecolumngrid
\copyright  2016 Optical Society of America. One print or electronic copy may be made for personal use only. Systematic reproduction and distribution, duplication of any material in this paper for a fee or for commercial purposes, or modifications of the content of this paper are prohibited.
\twocolumngrid

\begin{abstract}
The rapid generation of non-classical light serves as the foundation for exploring quantum optics and developing applications such as secure communication or generation of NOON-states. While strongly coupled quantum dot-photonic crystal resonator systems have great potential as non-classical light sources due to their promise of tailored output statistics, the generation of indistinguishable photons has been obscured due to the strongly dissipative nature of such systems. Here, we demonstrate that the recently discovered self-homodyne suppression technique can be used to overcome this limitation and tune the quantum statistics of transmitted light, achieving indistinguishable photon emission competitive with state-of-the-art metrics. Furthermore, our nanocavity-based platform directly lends itself to scalable on-chip architectures for quantum information.
\end{abstract}

\maketitle

Understanding the interaction between light and matter is of paramount importance for exploring the peculiar properties of quantum optics and utilizing them for applications such as non-classical light generation with implications for communication, information processing and sensing \cite{Gisin2002, Gisin2007, OBrien2007, Boto2000, Giovannetti2004}. In the solid-state, self-assembled quantum dots (QDs) are widely used as quantum emitters due to their strong interaction with light and ability to be integrated into nanophotonic resonators for enhanced light-matter interaction. Examples of quantum optical landmark experiments with QDs are the generation of indistinguishable photons \cite{Santori2002, Gazzano2013, He2013, Wei2014, Gschrey2015, Thoma2015},  entangled photon pairs \cite{Benson2000, Akopian2006,Trotta2014, Muller2014} and the observation of Mollow triplets \cite{Vamivakas2009, Flagg2009}.

For off-chip applications a high photon-extraction efficiency is desirable, which can be achieved by embedding the QD into a resonator with strong vertical emission. For example, QDs have been embedded in micropillar cavities \cite{Santori2002, Gazzano2013} that also Purcell enhance the emission rate for fast photon extraction. Utilizing resonant excitation of such structures, highly indistinguishable photon generation has recently been demonstrated \cite{Somasch2015, Ding2016, Unsleber2015}. Importantly, resonant excitation enables a high degree of indistinguishability due to the absence of excitation timing jitter and electric field noise that typically results from charge fluctuations in the semiconductor environment under non-resonant excitation \cite{He2013, Wei2014}.

On the other hand, for on-chip applications photonic crystal resonators are promising. Their planar geometry naturally allows for coupling with on-chip structures, including waveguides and on-chip detectors \cite{Reithmaier2013, Reithmaier2015} and hence holds promise to realize fully-integrated quantum optical hardware. Photonic crystal resonators provide extremely small mode volumes which enable a large enhancement of the light-matter interaction strength with embedded quantum emitters \cite{Yoshie2004, Englund2005, Hennessy2007}. Importantly, in transmission geometries for on-chip resonant generation of non-classical light can not be achieved by resonantly exciting a QD weakly coupled to a resonator. Resonant excitation of a weakly coupled QD-resonator system predominantly leads to coherent scattering from the cavity. Thus, in off-chip applications with micropillars, suppressing this coherent scattering while collecting emission from the QD using cross-polarized suppression is only enabled by a careful choice of quantum emitter structure. Specifically, the structure must allow that the polarization of the coherently scattered light is not rotated while the polarization of the QD emission is rotated. This can be achieved, for example, in bi-modal cavities and charge neutral QDs that have their symmetry axis different from the cavity and laser \cite{Giesz2015} or bi-modal cavities and charged QDs. Instead in photonic crystals, direct transmission of light through a strongly coupled QD-nanocavity system can generate a range of output quantum statistics \cite{Faraon2008, Reinhard2011, Laussy2014, Kai2015, Kai2015PRX}. However, the strongly dissipative nature of such systems has so far obscured the generation of indistinguishable photons.

In this letter, we demonstrate that interference which is intrinsic to photonic crystal cavities can be used to overcome this strongly dissipative nature and tune quantum statistics. Specifically, we show that this recently-discovered self-homodyne suppression (SHS) effect \cite{Kevin2015} can be used to interferometrically reject the coherent scattering off a dissipative Jaynes-Cummings system and isolate the non-classical component of the emitted light. While our experimental approach is tailored to photonic crystal cavities, self-homodyne suppression as a tool for engineering quantum statistics is widely applicable to other Jaynes-Cummings systems. Here, we demonstrate the robust and ultrafast generation of highly-indistinguishable photons from strongly coupled quantum dot-photonic crystal resonator systems with state-of-the-art indistinguishability and generation rates. Additionally, this approach circumvents the temperature limit, set by phonon-dephasing, in all previous solid-state approaches \cite{Santori2002, Gschrey2015, Thoma2015}  while also facilitating on-chip integration. 

\begin{figure}[!t]
\includegraphics[width=\linewidth]{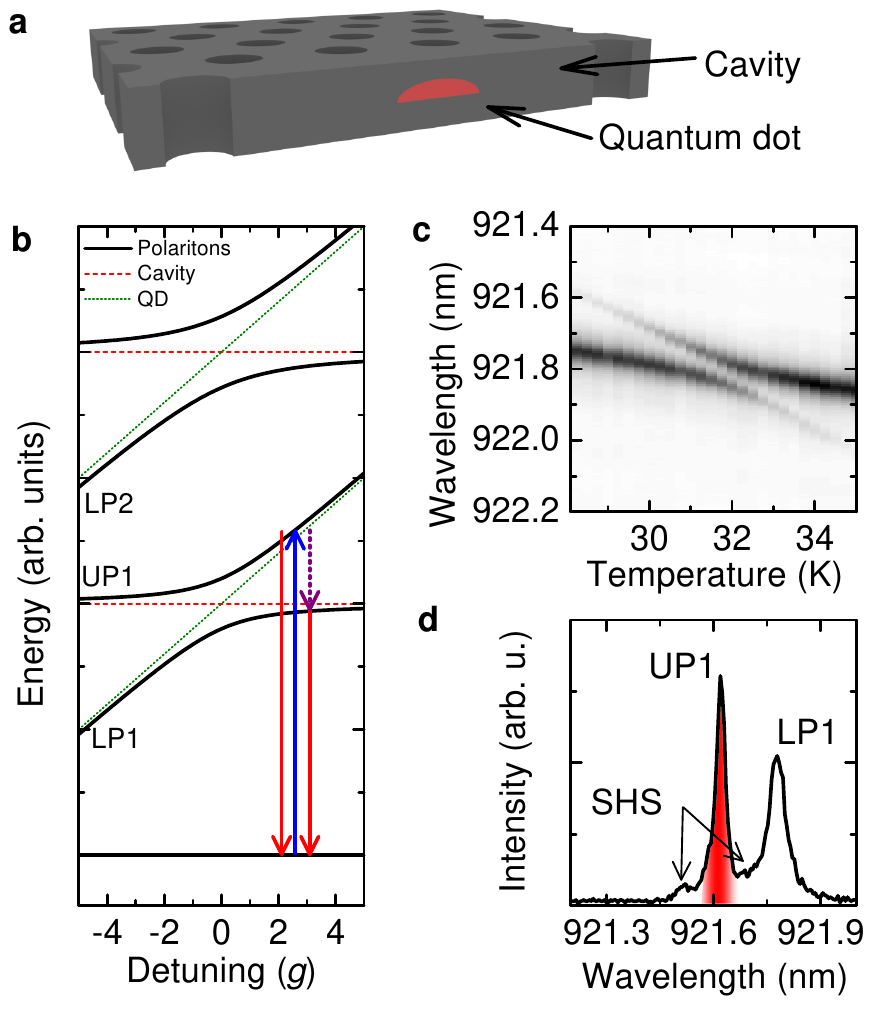}
\caption{\textbf{Resonantly excited strongly coupled system.} \textbf{a}, Schematic illustration of the QD-photonic crystal cavity platform. \textbf{b}, Schematic illustration of the Jaynes-Cummings ladder of dressed states that describes the energy level structure of a strongly coupled system. The arrows illustrate the resonant excitation of UP1 and subsequent relaxation. \textbf{c}, Cross-polarized reflectivity spectrum of the coupled quantum dot-cavity system obtained by temperature tuning the QD through the cavity resonance. An anticrossing of the peaks clearly demonstrates the strong coupling. \textbf{d}, Typical spectrum for resonantly exciting UP1 with a 16 ps long pulse at a QD-cavity detuning of $\Delta = 4.5\,g$ and in the presence of self-homodyne suppression. The red shaded region indicates the spectral filter used in subsequent experiments.}
\label{figure:1}
\end{figure}

The system under investigation consists of a single self-assembled quantum dot (QD) strongly coupled to a photonic crystal L3 cavity (Fig. \ref{figure:1}a). The resulting energy level structure is well described by the Jaynes-Cummings (JC) dressed states ladder. The energies of the lowest two rungs are presented in Fig. \ref{figure:1}b as a function of the QD-cavity detuning $\Delta$. They form pairs of anticrossing lines, labelled UP$n$ and LP$n$ for the upper and lower polariton of rung $n$, respectively. Experimentally this coupling can be observed in detuning-dependent cross-polarized reflectivity measurements (Fig. \ref{figure:1}c) that reveal the clear anticrossing of the first rung \cite{Englund2007}. As the polaritonic peaks transition through the avoided-crossing (at $\Delta$ = 0), they exchange character from cavity/QD-like to QD/cavity-like. Fitting the data results in a QD-cavity coupling strength of $g= 2\pi \cdot 12.3\,\textrm{GHz}$ and a cavity energy decay rate of $\kappa = 2\pi \cdot 18.4\,\textrm{GHz}$.

Resonant generation of single photons in such systems can be achieved by photon blockade \cite{Faraon2008}. Here, an excitation laser tuned in resonance with the first rung is out of resonance with higher climbs up the ladder due to the JC anharmonicity. However, these resonances have broad linewidths and hence appreciable overlap due to the highly dissipative character of semiconducting systems. Nevertheless, by detuning the QD and cavity by a few $g$ and exciting the QD-like polariton branch, high purity and efficiency single-photon generation has recently been demonstrated (blue arrow in Fig. \ref{figure:1}b) \cite{Kai2015}. To achieve the strongest photon blockade in pulsed on-demand applications, the pulse length has to be chosen as a compromise that minimizes both re-excitation and resonance overlap \cite{Kai2015PRX}. 

We now discuss photon blockade in the context of self-homodyne suppression. Here, the suppression results from destructive interference between the light scattered from the fundamental cavity mode and the continuum above-the-light-line modes, leading to Fano-like resonances \cite{Vasco2013}. This effect can be used to significantly suppress JC coherent scattering of the excitation laser in detuned QD-cavity systems and extract the incoherent spectrum \cite{Kevin2015}. To reach SHS, we optimized the excitation conditions (focus and polarisation) for the suppression of coherent scattering. The pulse length is chosen to be only $16\,\textrm{ps}$ to minimize re-excitation. The resulting spectrum (Fig. 1d) exhibits three distinct features: emission from the resonantly excited QD-like UP1, phonon-assisted emission from the cavity-like LP1 (purple and red arrows in Fig. 1b) and self-homodyne suppressed JC coherently scattered laser light. The last one is strongest on the sides of the UP1 peak due to a spectral dependence of SHS that results from the wavelength-dependent phase shift of the JC coherently scattered light.

To investigate the single-photon generation and photon indistinguishability under resonant excitation of UP1 (Fig. \ref{figure:1}d), we measure photon correlations between the outputs of a fiber-based Mach-Zehnder (MZ) interferometer (Fig. \ref{figure:2}a). Here, we excite the system with double pulses that each have a pulse area of $\pi$ (see Supplementary information) and a time delay $T_1=1.9\,\textrm{ns}$ that matches the delay of the interferometer. First, we perform experiments without spectral filtering. The result (Fig. \ref{figure:2}b) is a pattern of five peaks separated by $T_1$ and repeated with the repetition rate of the laser ($80\,\textrm{MHz}$). Due to the quantum character of the emission, the three centre peaks around zero time delay are attenuated \cite{Santori2002}. Note that the asymmetry of the five peaks results from the imperfect reflectivity to transmittivity ratios of the second beamsplitter in the fiber-coupled implementation (see Supplementary information for details). To quantitatively analyze the data we bin the counts in a time window of $384\,\textrm{ps}$ about the peaks (Fig. 2b datapoints). A fit to the data (blue columns in Fig. \ref{figure:2}c) allows for extraction of the measured degree of second-order coherence $g^{(2)}[0]$ and first-order coherence $|g^{(1)}[0]|$ between two subsequent pulses. The extracted values of the fit are $g^{(2)}[0]=0.24 \pm 0.03$ and $|g^{(1)}[0]|^2=0.25 \pm 0.03$. In the literature, when analyzing the attenuation of the center peak instead of $|g^{(1)}[0]|^2$, a quantity $v$ is often stated and defined as the single-photon mode overlap. However, this parameter $v$ would only correspond to the single-photon mode overlap for pulses of perfect single-photon character ($g^{(2)}[0]=0$). The limited fidelity of the measurement can be understood by recalling the emission spectrum presented in Fig. \ref{figure:1}d. The imperfect suppression of the JC coherently scattered light limits $g^{(2)}[0]$, while the phonon-assisted emission from LP1 limits $|g^{(1)}[0]|^2$ due to excitation timing jitter.

\begin{figure}[!t]
\includegraphics[width=\linewidth]{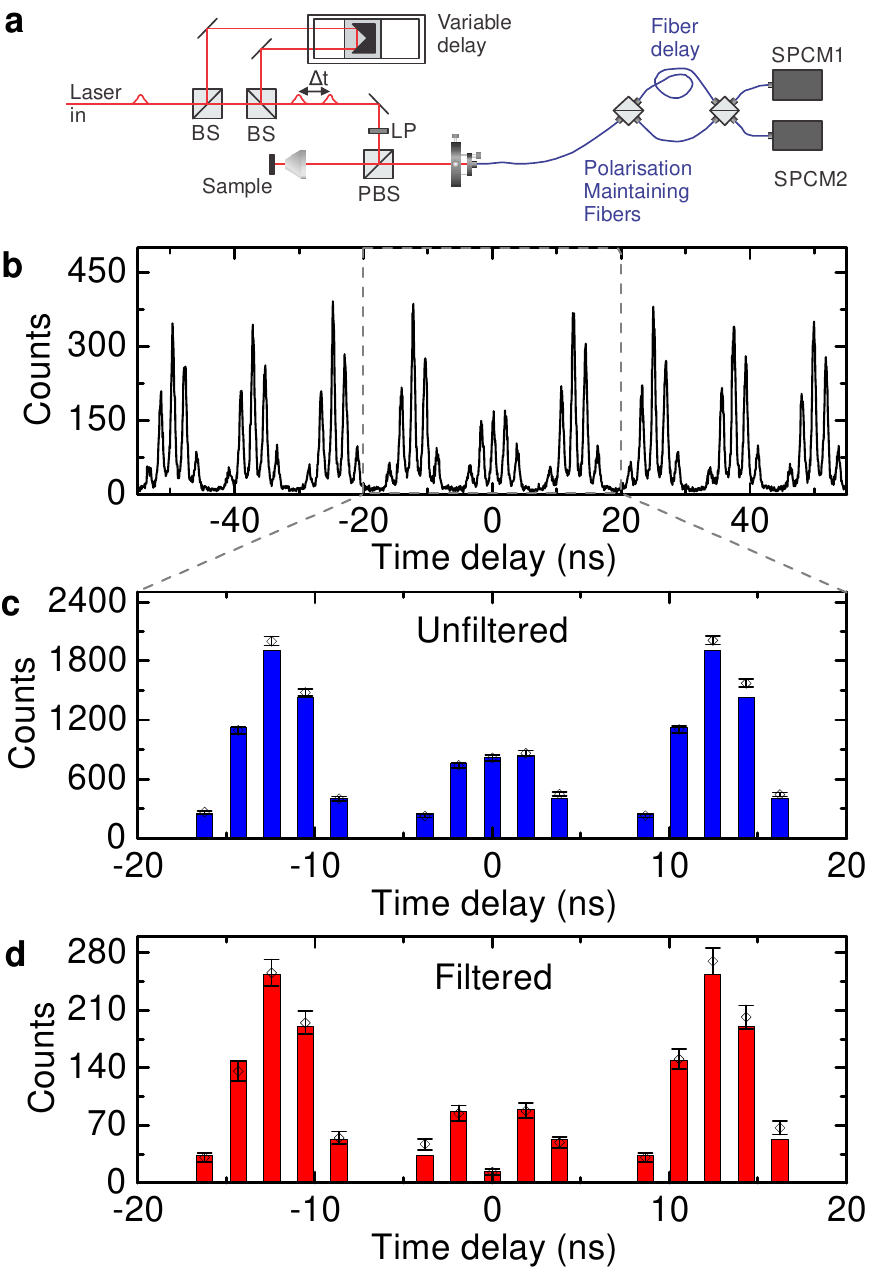}
\caption{\textbf{Indistinguishability measurements.} \textbf{a}, Schematic illustration of the setup used to extract Hong-Ou-Mandel interference. \textbf{b}, Measured correlation function of the emission using the same excitation conditions as in Fig. \ref{figure:1}d. Due to the quantum character of the light, the amplitude of the three centre peaks surrounding zero time delay is reduced. \textbf{c}, Amplitudes around zero delay obtained from binning the data presented in \textbf{b} with a temporal width of $384\,\textrm{ps}$ about the centre of each peak (represented as diamond datapoints). The error bars result from the $\sqrt{N}$ variation of the photocount distribution. Fits to the data are presented as blue columns and reveal $g^{(2)}[0]=0.24 \pm 0.03$ and $|g^{(1)}[0]|^2=0.25 \pm 0.03$. \textbf{d}, Same as \textbf{c} but under spectral filtering of the emission from UP1, resulting in $g^{(2)}[0]=0.05\pm0.04$ and $|g^{(1)}[0]|^2=0.96\pm0.05$}
\label{figure:2}
\end{figure}

To increase the fidelity of indistinguishable photon generation through photon blockade, we now employ spectral filtering. Therefore, we repeat the correlation measurement while filtering on the UP1 emission, as indicated by the red shaded region in Fig. \ref{figure:1}d. The result of this experiment with a filter bandwidth of $2 \pi \cdot 10\,\text{GHz}$ is presented in Fig. \ref{figure:2}d as empty diamonds. A fit (red columns) extracts values of $g^{(2)}[0]=0.05\pm 0.04$ and $|g^{(1)}[0]|^2=0.96 \pm 0.05$. Note that this bandwidth is much larger than the linewidth of the UP1 emission. Therefore, the improvements result only from eliminating imperfect suppression of JC coherently scattered light and phonon-assisted emission from LP1 and not from filtering with a bandwidth smaller than spectral diffusion of the QD. In both cases, frequency filtered and unfiltered we confirmed the extracted values of $g^{(2)}[0]$ in second-order correlation measurements and obtained similar values. Our metrics are competitive with the best values obtained from QDs so far \cite{He2013, Wei2014, Somasch2015, Ding2016}. The UP1 state lifetime in our experiment of $55\,\textrm{ps}$ (measured at this detuning \cite{Kai2015PRX}) paves the way for on-chip generation rates over an order of magnitude faster than bulk QDs and slightly faster than those in micropillar resonators \cite{Somasch2015, Ding2016, Unsleber2015}. However, micropillar resonators are optimized for photon extraction leading to higher count rates and they also do not require spectral filtering. Nevertheless, as discussed above the photonic crystal platform facilitates scalable on-chip architectures and in this platform, its near unity coupling efficiency to waveguides matters over its emission profile. Finally, the measurements presented here have been performed at a relatively high temperature of approximately 30 K, directly contrasting with the best previously reported HOM interference visibility of $<40\%$ at such a temperature \cite{Thoma2015}. This difference can be understood from interaction with the high-temperature phonon bath: In bulk, the interaction with phonons results in dephasing of the emission which reduces the first-order coherence. Meanwhile in a strongly coupled system, the interaction with phonons leads to a population transfer from UP1 to LP1 \cite{Kai2015PRX}, spectrally removing the dephased emission from the detection channel and ensuring robust high-fidelity operation. Thus, we have investigated indistinguishable photon generation in a dephasing regime unlike all previous experiments and found that this region is highly beneficial for photon indistinguishability.

To corroborate our finding that the combination of SHS and spectral filtering results in high-fidelity generation of indistinguishable photons we performed quantum optical simulations (see  Supplementary information). The simulated emission spectrum using the measured system parameters and excitation with a $\pi$ pulse is presented in Fig. \ref{figure:3}a without (blue) and with optimized (black) SHS (see Supplementary information for details). Only when including SHS is the spectrum in extremely good qualitative agreement with the experimentally measured one. To demonstrate the impact of SHS on the single-photon character of the emission, we simulate $g^{(2)}[0]$ as a function of the SHS strength without (Fig. \ref{figure:3}b black) and with (Fig. \ref{figure:3}b red) spectral filtering of the emission from UP1 . Here, the parameter $\alpha$ denotes the intensity of the continuum-mode scattered contribution. In both cases, a clear dip in values of $g^{(2)}[0]$ very close to the measured values is obtained, with much lower $g^{(2)}[0]$ for the filtered system. Moreover, at the point of best suppression, the simulations reveal that the average number of photons exiting the system per pulse is unity. 

\begin{figure}[!t]
\includegraphics[width=\linewidth]{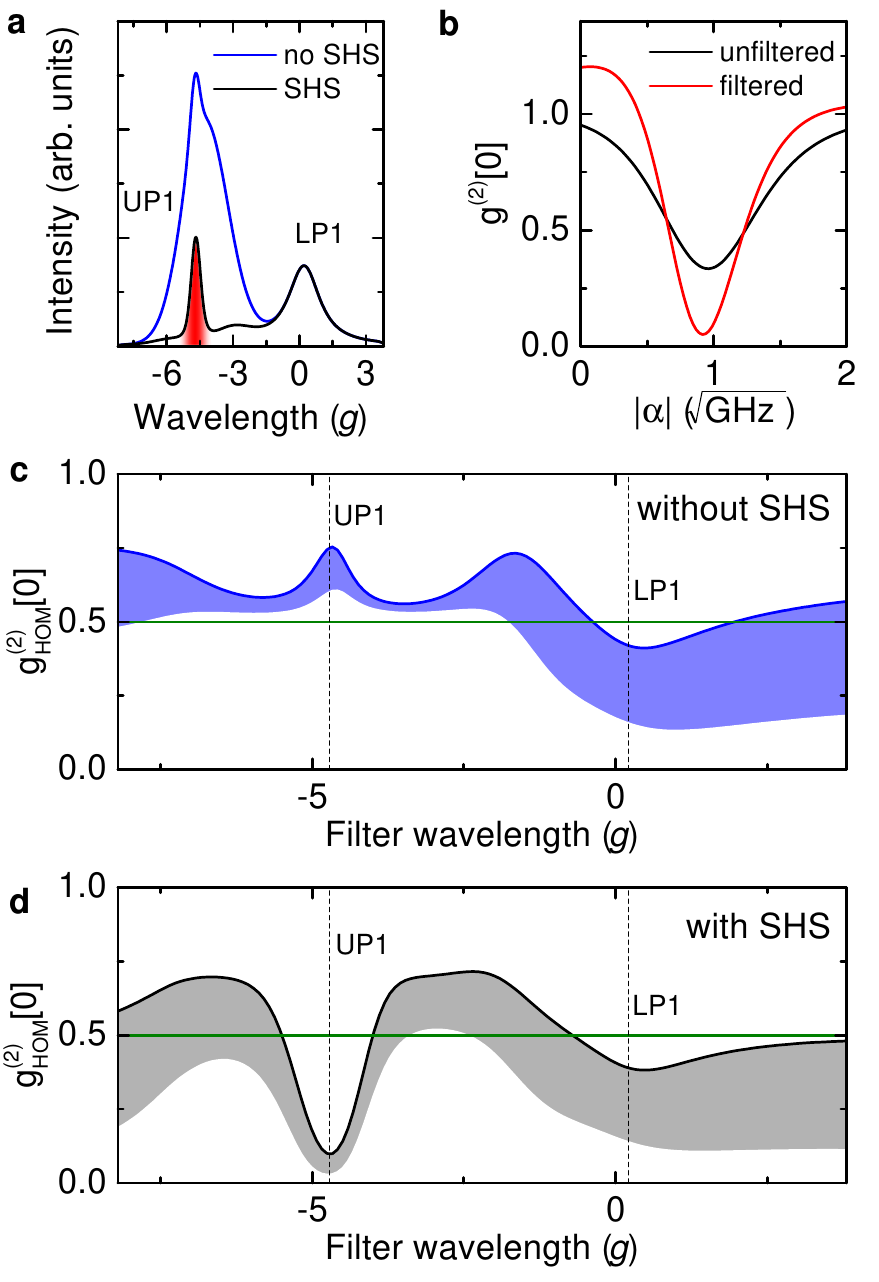}
\caption{\textbf{Quantum optical simulations.} \textbf{a}, Simulated spectrum for resonant excitation of UP1 by a $16\,\textrm{ps}$ long $\pi$ pulse at a QD-cavity detuning  $\Delta = 4.5\,g$, with and without SHS. \textbf{b}, $g^{(2)}[0]$ (second-order coherence) for the excitation conditions of \textbf{a} (resonant excitation of UP1) as a function of the SHS tuning parameter with and without spectral filtering on UP1 (red region in \textbf{a}). \textbf{c-d}, $g^{(2)}_{HOM}[0]$ as a function of the filtering wavelength excluding (\textbf{c}) and including (\textbf{d}) SHS. Only the central wavelength of the filer changes, but bandwidth remains the same. The shaded area below the curves visualizes $\frac{1}{2}(1-|g^{(1)}[0]|^2)$ while the white area visualizes $\frac{1}{2} g^{(2)}[0]$. Green lines denote the nonclassical threshold. All wavelengths are referenced to that of the bare cavity.}
\label{figure:3}
\end{figure}

To investigate photon indistinguishability, we calculate the second-order cross correlation $g^{(2)}_{HOM}[0]$  discussed in the original Hong-Ou-Mandel (HOM) paper \cite{HOM1987}. Here, correlations between the output ports of a beamsplitter are simulated while two identical systems feed the input ports. It is important to note that in contrast to the often-confused statements in the literature, in this configuration the dip in the correlation $g^{(2)}_{HOM}[0]$ is different than the dip of the center peak of the Fig. \ref{figure:2}a MZ implementation (see Supplementary information for details). This distinction is important because the MZ scheme is now predominantly used to experimentally characterize single-photon source indistinguishability in our and other experiments. Here, an important result of our work is that the HOM configuration shows a dip of
\begin{equation}
g^{(2)}_{HOM}[0] = \frac{1}{2} g^{(2)}[0] + \frac{1}{2} \left[1-|g^{(1)}[0]|^2\right]
\end{equation}
while the MZ zero delay dip reduces to:
\begin{equation}
g^{(2)}_{MZ}[0] = \frac{2}{3} g^{(2)}[0] + \frac{1}{3} \left[1-|g^{(1)}[0]|^2\right]
\end{equation}
Therefore, the two are equal only for $g^{(2)}[0]=0$ and $g^{(1)}[0]=1$. To visualize this difference, we look at our experimentally obtained values of $g^{(2)}[0]$ and $|g^{(1)}[0]|^2$ discussed above. For the frequency filtered case we obtain $g^{(2)}_{HOM}[0]=0.045\pm0.045$ and $g^{(2)}_{MZ}[0]=0.047\pm0.043$ while for the unfiltered case we obtain $g^{(2)}_{HOM}[0]=0.495\pm0.03$ and $g^{(2)}_{MZ}[0]=0.41\pm0.03$. In the filtered case, the difference is small due to the comparable values of $g^{(2)}[0]$ and $1- |g^{(1)}[0]|^2$ but in the unfiltered case the difference is significant. Nevertheless, when extracting $g^{(2)}[0]$ and $|g^{(1)}[0]|^2$, it is possible to directly compare the values obtained from the two different methods, as well as simulations and experiment. 

The results of the HOM simulations are presented in Figs \ref{figure:3}c and \ref{figure:3}d excluding and including SHS, respectively. The figures show $g^{(2)}_{HOM}[0]$ against the filtering wavelength, as solid lines. The area below the line is decomposed into $\frac{1}{2} (1-|g^{(1)}[0]|^2)$ (shaded) and $\frac{1}{2} g^{(2)}[0] $ (white). Without SHS only a weak dip at the wavelength of LP1 is observed. However, it is barely nonclassical since at this wavelength the emission is phonon-mediated and, thus, subject to strong excitation timing jitter which is reflected in the low value of $|g^{(1)}[0]|^2$. In contrast, when including SHS, a strong dip at the filtered wavelength of UP1 is observed with values of $g^{(2)}[0] = 0.05$ and $|g^{(1)}[0]|^2 = 0.86$, in excellent agreement with the experimentally measured values. 

In summary, we have demonstrated that the self-homodyne technique can be used to interferometrically tune the output quantum statistics of Jaynes-Cummings systems. In particular, we showed that strongly coupled QD-photonic crystal nanocavity systems are capable of robust and high-fidelity generation of indistinguishable photons even at elevated temperatures, by combining resonant excitation, self-homodyne suppression and spectral filtering. Having produced indistinguishable photons from a state with a lifetime of only $55\,\textrm{ps}$, our results could pave the way for sources with unprecedented rates. Moreover, the short lifetime leads to a homogeneous linewidth of the emission which is  much larger than that of bulk QDs. Therefore, we expect the indistinguishability to be unaffacted by spectral diffusion of the quantum emitter even without active suppression of the spectral diffusion \cite{Prechtel2013, Hansom2014, Kuhlmann2015}. Specifically, in contrast to bulk QDs \cite{He2013, Thoma2015} we expect a similarly high indistinguishability for excitation pulses with a longer time delay or for measurements interfering the emission from multiple systems. Furthermore, the generation of indistinguishable photons from strongly coupled QD-photonic crystal systems enables scalable on-chip architectures. While in our case the positioning of the QD relative to the cavity has been done probabilistically, recent progress in site-selective growth of QDs \cite{Yakes2013, Unsleber2015-2} as well as positioning of resonators relative to QDs \cite{Somasch2015, Gschrey2015, Unsleber2015} provides further support for the feasibility of integrated quantum photonic circuits. Finally, the demonstration of the self-homodyne technique to isolate the quantum character of resonantly scattered light paves the way for photon bundling \cite{Laussy2014} at significantly lower nonlinearity-cavity detunings and smaller powers for high-throughput generation of other nonclassical states of light.

\section*{Funding Information}
We gratefully acknowledge financial support from the Air Force Office of Scientific Research, MURI center for multifunctional light-matter interfaces based on atoms and solids (Grant No. FA9550-12-1-0025), support from the Army Research Office (Grant No. W911NF1310309) and support from the National Science Foundation (Division of Materials Research - Grant. No. 1503759). KM acknowledges support from the Alexander von Humboldt Foundation. KAF acknowledges support from the Lu Stanford Graduate Fellowship and the National Defense Science and Engineering Graduate Fellowship. CD acknowledges support from the Stanford Graduate Fellowship. KGL acknowledges support from the Swiss National Science Foundation. YAK acknowledges support from the Stanford Graduate Fellowship and the National Defense Science and Engineering Graduate Fellowship.


\providecommand{\noopsort}[1]{}\providecommand{\singleletter}[1]{#1}%

\cleardoublepage

\onecolumngrid
\section*{Supplemental Material}
$ $\\
$ $\\
$ $\\
\twocolumngrid

\section*{Methods}
\textbf{Sample fabrication:} The MBE-grown structure consists of an $\approx 900\,\textrm{nm}$ thick $\textrm{Al}_{0.8} \textrm{Ga}_{0.2} \textrm{As}$ sacrificial layer followed by a $145\,\textrm{nm}$ thick GaAs layer that contains a single layer of InAs QDs. Our growth conditions result in a typical QD density of $(60 - 80)\,\mu\textrm{m}^{-2}$. Using $100\,\textrm{keV}$ e-beam lithography with ZEP resist, followed by reactive ion etching and HF removal of the sacrificial layer, we define the photonic crystal cavity. The photonic crystal lattice constant was $a = 246\,\textrm{nm}$ and the hole radius $r \approx 60 \,\textrm{nm}$. The cavity fabricated is a linear three-hole defect (L3) cavity with a measured quality factor of $Q=17700$. To improve the cavity quality factor, holes adjacent to the cavity were shifted \cite{Akahane2005, Minkov2014}.

\textbf{Optical spectroscopy:} All optical measurements were performed with a liquid helium flow cryostat at temperatures in the range $10-40 \, \text{K}$. For excitation and detection a microscope objective with a numeric aperture of $NA = 0.75$ was used. Cross-polarised measurements were performed using a polarising beam splitter. To further enhance the extinction ratio, additional thin film linear polarisers were placed in the excitation/detection pathways and a single mode fibre was used to spatially filter the detection signal. Furthermore, two waveplates were placed between the beamsplitter and microscope objective: a half-wave plate to rotate the polarisation relative to the cavity and a quarter-wave plate to correct for birefringence of the optics and sample itself. We estimate our overall efficiency to be $\sim 0.125\%$, including setup losses and detector efficiency. Excluding the detector efficiency this corresponds to a collection efficiency into the detection single mode fiber of $\sim 0.5\%$

\textbf{Correlation measurements:} Second-order autocorrelation measurements were performed using a Hanbury Brown and Twiss (HBT) setup consisting of one fibre beamsplitter and two single-photon avalanche diodes. The  detected photons were correlated with a PicoHarp300 time counting module. Two-photon interference measurements were performed with the same detectors and electronics as the second-order correlation measurements but with a Mach-Zehnder interferometer replacing the single beamsplitter, as discussed in the main text. For the unfiltered two-photon interference measurements the integration time was 30 minutes and the coincidence rate (in the whole histogram) was 2336/s. For the frequency filtered two-photon interference measurements the integration time was 8.6 hours and the coincidence rate (in the whole histogram) was 22/s.

\section*{\textsf{Details on the simulations}}
The quantum-optical simulations were performed using density matrix master equations with the Quantum Optics Toolbox in Python (QuTiP)\cite{Johansson2013}, where the standard\cite{Alsing1992} Jaynes-Cummings model was used as a starting point. The effects of phonons were incorporated through the addition of incoherent decay channels with rates that were previously extracted\cite{Kai2015PRX}.

To simulate the first order spectra of our system under excitation with a single pulse, we compute the one-sided spectrum
\begin{equation} \label{eq:1}
S(\omega)=\textrm{Re} \left[ \iint_{\mathbb{R}^2} \mathop{\textrm{d} t} \mathop{\textrm{d} \tau} \textrm{e}^{-\textrm{i}\omega\tau} \langle A^\dag(t+\tau) A(t) \rangle \right]
\end{equation}
of the free-field mode operator $A(t)$. Input-output theory can relate the internal cavity mode operator $a(t)$ to the external field operator by the radiative cavity field decay rate $\kappa/2$. Hence, for a JC system in the solid state where the QD radiative decay rate $\gamma$ plays an insignificant role compared to $\kappa$\cite{Kai2015PRX}, spectral decomposition of the cavity mode operator yields the spectrum of the detected light. Therefore, we can compute an unnormalised version of this spectrum with $A(t) \rightarrow a(t) $ in equation (1). We can also compute an unnormalised version of the incoherent spectrum with $\langle A^\dag(t+\tau) A(t) \rangle \rightarrow \langle A^\dag(t+\tau) A(t) \rangle - \langle A^\dag(t+\tau) \rangle \langle A(t) \rangle$ in equation (1). To arrive at the version measured by a spectrometer of finite bandwidth, we convolve $S(\omega)$ with the spectrometer's response function. To simulate self-homodyne suppression (SHS), we replace $A(t) \rightarrow a(t)+\alpha(t)$ in equation (1). Physically, $\alpha(t)$ is a slightly phase- and amplitude- shifted version of the incident laser pulse (originating from the continuum-mode scattering)\cite{Kevin2015}.

In order to simulate the normalised measured degree of second-order coherence, $g^{(2)}[0]=\frac{G^{(2)}[0]}{N^2}$ with ${N=\int_{\mathbb{R}} \mathop{\textrm{d} t} \langle A^\dag(t) A(t) \rangle}$, we calculate
\begin{equation} \label{eq:2}
g^{(2)}[0]=\frac{\iint_{\mathbb{R}^2} \mathop{\textrm{d} t} \mathop{\textrm{d} \tau} \langle \mathcal{T}_- [A^\dag(t) A^\dag(t+\tau)] \mathcal{T}_+ [A(t+\tau) A(t)] \rangle}{\left(\int_{\mathbb{R}} \mathop{\textrm{d} t} \langle A^\dag(t) A(t) \rangle \right)^2}
\end{equation}
under excitation by a single pulse\cite{Kai2015}. The operators $\mathcal{T}_\pm$ indicate the time ordering required of a physical measurement\cite{Elena2012}. We can likewise replace $A(t) \rightarrow a(t)$ in equation (2) and also model SHS with the replacement of $A(t) \rightarrow a(t)+\alpha(t)$ in equation (2). Despite the simplicity of equation (1), adding spectral filtering to equation (2) is analytically and numerically quite challenging. The spectral decomposition of this equation requires a fourth order integral that is often intractable even numerically. Fortunately, the newly discovered sensor formalism\cite{Elena2012} allows for efficient calculation of the spectrally filtered version of the measured degree of second-order coherence. Here, we coherently attach a pair of two-level sensors to the system Hamiltonian with the addition of the sensor Hamiltonian to the Jaynes-Cummings Hamiltonian:
\begin{equation} \label{eq:3}
H = H_\textrm{JC} + \sum_{i=1}^2\left[ \omega_s \varsigma^\dag_i \varsigma_i + 
\epsilon\left( a \varsigma_i^\dag + a^\dag \varsigma_i \right) \right]
\end{equation}
where $\omega_s$ is the sensor frequency, $\varsigma$ the sensor annihilation operator, and $\epsilon$ the sensor coherent coupling strength. The sensor coupling is chosen small enough so that its backaction on the system is negligible, i.e. $\frac{\epsilon^2}{\Gamma/2} \ll \gamma_f$ where $\gamma_f$ is the fastest transition rate in the un-sensed system. Additionally, the sensor decay terms of rate $\Gamma$ are added to the total Liouvillian. Here, in order to simulate SHS, we replace
\begin{equation} \label{eq:4}
a \varsigma_i^\dag + a^\dag \varsigma_i \rightarrow \left(a+\left<\alpha(t)\right>\right)\cdot\varsigma_i^\dag + \left(a^\dag+\left<\alpha^{*} (t)\right>\right) \cdot \varsigma_i
\end{equation}
in equation (3).

To arrive at the physically measured and spectrally filtered second-order coherence functions, the total degree of second-order coherence is computed between the two sensors:
\begin{equation} \label{eq:5}
g^{(2)}[0]=\frac{\iint_{\mathbb{R}^2} \mathop{\textrm{d} t} \mathop{\textrm{d} \tau} \langle \mathcal{T}_- [\varsigma_1^\dag(t) \varsigma_2^\dag(t+\tau)] \mathcal{T}_+ [\varsigma_1(t+\tau) \varsigma_2(t)] \rangle}{\left(\int_{\mathbb{R}} \mathop{\textrm{d} t} \langle \varsigma_1^\dag(t) \varsigma_1(t) \rangle\right)^2}
\end{equation}
As the sensors are degenerate in every manner, the ordering of their operation is arbitrary. In our model, the sensors are used as filters while the detector is assumed to be sufficiently broad-band to integrate the correlations over our entire experimental domain. This approximation is accurate as the detector has a timing resolution of greater than $200\,\textrm{ps}$ compared with the system decay time of approximately $65\,\textrm{ps}$. Additionally, we integrate our data over $\approx 400\,\textrm{ps}$ bins.

\section*{\textsf{Simulated power-dependent spectra}}
\begin{figure}[!t]
\includegraphics[width=\columnwidth]{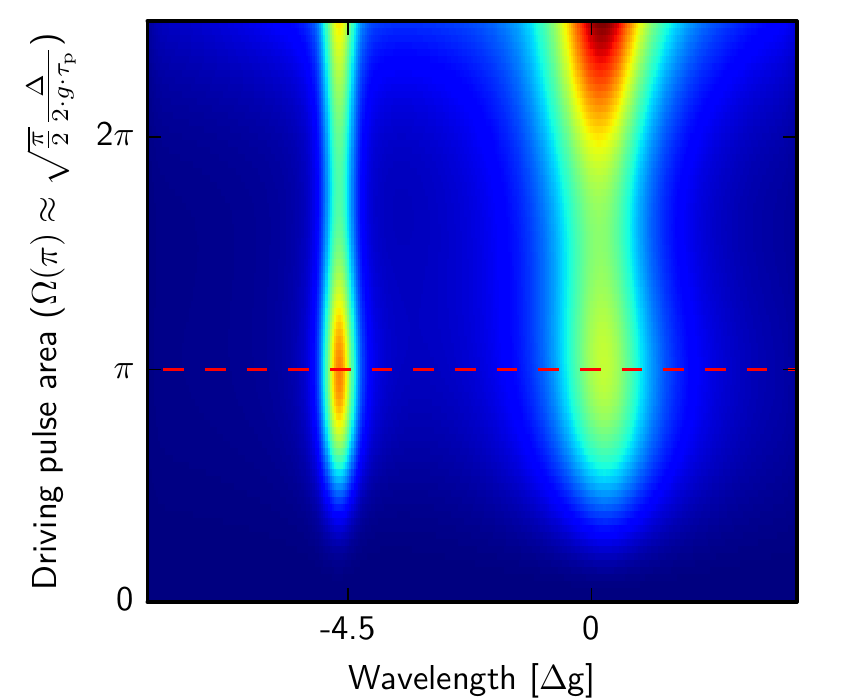}
\caption{\textbf{Incoherent spectra.} Power dependent spectra for resonantly exciting UP1 with $16\,\textsf{ps}$ short pulses at a QD-cavity detuning of $\Delta = 4.5\,g$.}
\label{figure:S1}
\end{figure}
We began our simulations by computing the incoherent power spectrum to find the $\pi$ pulse driving power. As all experiments were performed with a $\pi$ pulse driving, we wanted to ensure an accurate match with simulations. Since the Rabi oscillations are clearly present in the incoherent spectrum without having to optimise SHS, this served as a simple way to verify the $\pi$ pulse power. The result of such a simulation is presented in Fig. \ref{figure:S1} for a resonant excitation of UP1 with $16\,\textrm{ps}$ short pulses at a QD-cavity detuning of $\Delta=4.5g$. For this detuning, the forward (UP1 to LP1) and reverse (LP1 to UP1) phonon-assisted transfer rates are $2\pi \cdot 2.4\,\textrm{GHz}$ and $2\pi \cdot 2.2\,\textrm{GHz}$ respectively.

 Note that after optimizing SHS, a nearly identical simulation of the measured spectrum is observed (see main text). For both UP1 and LP1 Rabi oscillations are observed and the pulse area for a $\pi$ pulse is close to $\sqrt{\frac{\pi}{2}}\frac{\Delta}{2 g \tau_p}$ where $\tau_p$ is the width parameter of the Gaussian pulse. This analytic result comes from truncating the detuned Jaynes-Cummings level structure and calculating the Raman transition frequency between the lowest-lying levels\cite{Kevin2015}. From there, the $\pi$ pulse driving strength is given approximately by integrating the pulse area - the exact drive area is difficult to analytically calculate due to the nonlinear damping of the system.

\section*{\textsf{Hong-Ou-Mandel interference}}

\begin{figure*}[!t]
\centering
\includegraphics[width=17cm]{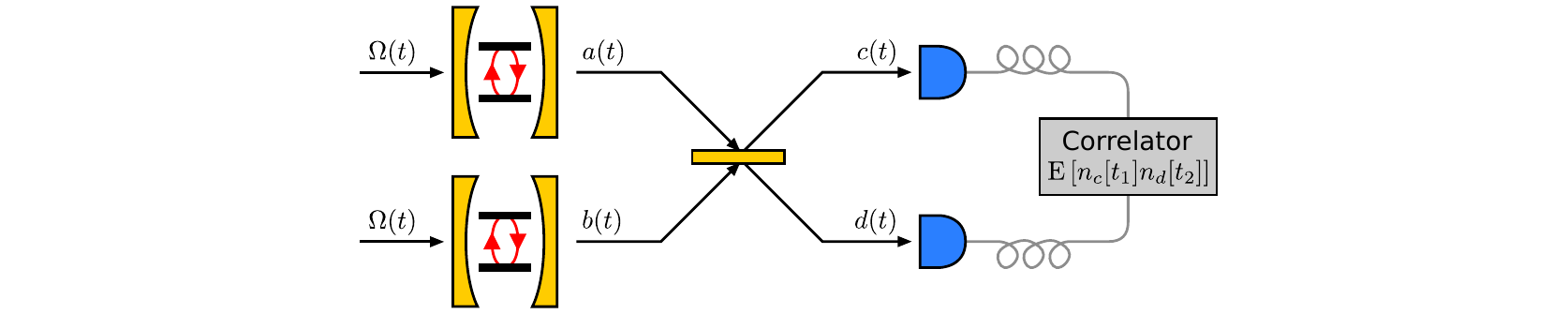}
\caption{\textbf{Schematic of the Hong-Ou-Mandel interferometer.} In this interferometer, two driven strongly coupled systems are interfered on two detectors by a beamsplitter. A digital recorder then correlates the detection times and computes $\textrm{E}[n_c[t_1]n_d[t_2]]$. Here, $\Omega(t)$ indicates the coherent driving field, $a(t)$ and $b(t)$ indicate the continuous mode free-field operators at the outputs of the driven systems, and $c(t)$ and $d(t)$ indicate the free-field operators at the inputs to the detectors.}
\label{figure:S2}
\end{figure*}
\begin{figure*}[!t]
\centering
\vspace{10pt}
\includegraphics[width=17cm]{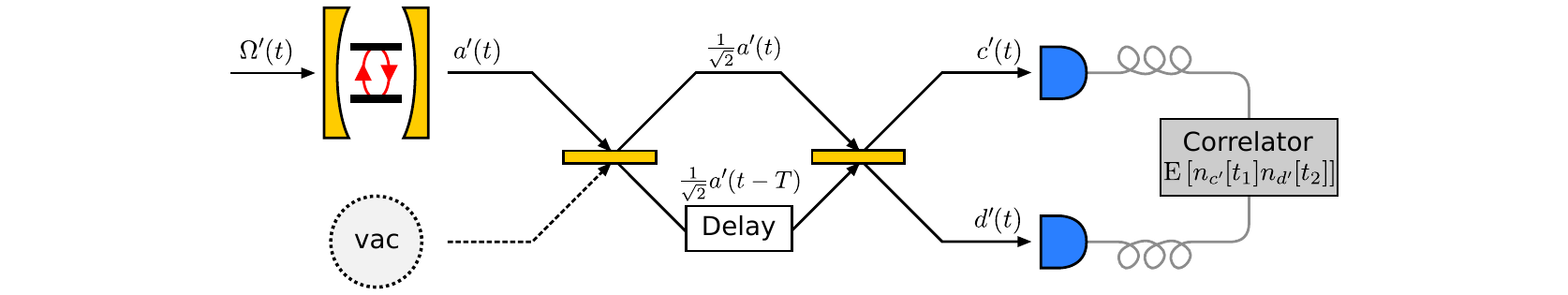}
\caption{\textbf{Schematic of the Mach-Zehnder interferometer.} In this interferometer, a single driven strongly coupled system is doubly excited at an interval of $T=1.9\,\textrm{ns}$, to be interfered with a time-delayed copy of itself. Importantly, the re-excitation only occurs after all excited population from the first pulse has decayed, so that the operator $a'(t)$ is identically independent between excitations. The vacuum mode operator at the input to the first beamsplitter has been omitted in anticipation that our detectors only measure a normally ordered moment of their input fields. A digital recorder then again correlates the detection times and computes $\textrm{E}[n_{c'}[t_1]n_{d'}[t_2]]$. Here, $\Omega'(t)=\Omega(t) + \Omega(t-T)$ indicates the coherent driving field, $a'(t)=a(t)+a(t-T)$ indicates the continuous mode free-field operator at the output of the driven system, and $c'(t)$ and $d'(t)$ indicate the free-field operators at the inputs to the detectors.}
\label{figure:S3}
\end{figure*}

We first review interference in a Hong-Ou-Mandel (HOM) interferometer. In such an interferometer, two identically independent sources are interfered on two detectors by the action of a single beamsplitter. This setup is shown schematically in Fig. \ref{figure:S2}. Here, we consider two strongly coupled systems each driven with the same pulse $\Omega(t)$. Their outputs, the Heisenberg free-field operators $a(t)$ and $b(t)$, are then fed into the beamsplitter which mixes the two according to the unitary transformation
\begin{equation} \label{eq:6}
\left[ \begin{array}{c}
c(t) \\
d(t) \\
\end{array} \right]
=
\frac{1}{\sqrt{2}}
\left[ \begin{array}{cc}
1 & -1 \\
1 & 1 \\
\end{array} \right]
\left[ \begin{array}{c}
a(t) \\
b(t) \\
\end{array} \right]
\end{equation}
The detection events are then correlated electronically to arrive at a temporal coincidence histogram $\textrm{E}[n_c[t_1]n_d[t_2]]$. In terms of the instantaneous correlations aforementioned, we wish to calculate $\textrm{E}[n_c[t_1]n_d[t_2]] = G^{(2)}_{cd}[0]$. In order to perform this calculation, we first decompose the underlying correlations of $G^{(2)}_{cd}\left(t_1, t_2 \right)$ into more manageable components based on $a(t)$ and $b(t)$ that will allow us to average over all quickly varying phase terms. The phase terms are difficult to observe experimentally since they depend on femtosecond phase locking and are, regardless, not required for the observation of two-photon interference. It has been shown that after performing this procedure\cite{Woolley2013-sz} one arrives at
\begin{multline} \label{eq:7}
G^{(2)}_{cd}\left(t_1, t_2 \right) = \frac{1}{2} G^{(2)}_a \left(t_1, t_2 \right) \\ + \frac{1}{2}\left[G^{(1)}_a \left(t_1, t_1 \right) \cdot G^{(1)}_a \left(t_2, t_2 \right)  - \left| G^{(1)}_a \left(t_1, t_2 \right) \right|^2 \right]
\end{multline}
where the correlations between $c(t)$ and $d(t)$ have terms only coming from a single source (the indices could be trivially switched as $a(t)$ and $b(t)$ are identically independent). Experimentally due to unknown detection efficiencies, our only true intensity reference is in proportion to the source correlation $N_a^2$. Combining this information with the discussion on detector bandwidth above, we are justified to compute only the integrated and normalised correlation
\begin{equation} \label{eq:8}
g^{(2)}_\textrm{HOM}[0] = \frac{1}{2} g^{(2)}_a [0] + \frac{1}{2} \left[1 - \left|g^{(1)}_a [0]\right|^2\right]
\end{equation}
referenced in the main text (equation [1]). Hence the observable HOM interference for indentically independent sources simply consists of an instantaneous measurement that depends on the total first- and second-order coherence of the source. Whether this interference is truly just two-photon interference therefore depends on the nature of the source output. Additionally, we note that in many real systems the first-order coherence extracted here is wildly different than the first-order coherence extracted from a Michelson interferometer due to the long-time averaging action of the Michelson interferometer. 

\section*{\textsf{Mach-Zehnder interference}}

Experimentally, producing a single source of indistinguishable photons is often quite challenging (let alone two), however we would still like to characterise the source's instantaneous degree of first order coherence. By doing so, we hope to quantify how it would behave in a HOM experiment. It was realised that some aspect of two-photon interference can be observed for a single source at the input to a Mach-Zehnder (MZ) interferometer\cite{Santori2002}. In order to observe this interference, the system must be doubly excited at a time interval that matches the temporal delay between the two paths of the MZ interferometer (shown schematically in Fig. \ref{figure:S3}). Additionally, this interval must be long enough that the emission resulting from the two excitations is identically independent. 

For perfect single photon input the MZ output correlation is in fact two-photon interference and thus equivalent to HOM interference. However, in may cases the literature has incorrectly compared the two when the source has some finite probability of emitting non-single photons. Thus, we now outline the derivation of the correct measured correlation, as discussed in the main text (equation [2]). Consider the schematic in Fig. \ref{figure:S3}: A single strongly coupled system is driven with the coherent driving field $\Omega'(t)=\Omega(t) + \Omega(t-T)$ and has field-mode operator output $a'(t)=a(t)+a(t-T)$. The first beamsplitter simply splits $a'(t)$ in two. Here we have dropped the vacuum operator in anticipation that our detectors only measure a normally ordered moment of their input fields. Next, one path is delayed by the excitation interval $T=1.9\,\textrm{ns}$. The second beamsplitter then mixes $\frac{1}{\sqrt{2}}a'(t)$ with the time delayed version of itself and the detection apparatus computes $\textrm{E}[n_{c'}[t_1]n_{d'}[t_2]] = G^{(2)}_{c'd'}[0]$ like before.

The time delay is absolutely critical: the purpose is that for time delays much less than $T$, $\frac{1}{\sqrt{2}}a'(t)$ can interfere with itself, and therefore reproduce a correlation similar to the HOM cross-correlation. Although computation of $g^{(2)}_{c'd'}[0]$ in terms of source correlations certainly appears daunting, the independence of $a'(t)$ for times greater than $T$ dramatically simplifies the calculation of $g^{(2)}_{c'd'}[0]$. This expansion is further simplified by consideration of the correlations centred around time delays that are integer multiples of $T$. It is fairly trivial to show that the majority of terms for delays larger than $T$ are zero, with only a few additional phase-dependent terms arising. We note that although one might expect the correlations to be phase locked by the MZ interferometer, our integration times of many hours quickly destroy the phase interference. Since these terms once again play no role in two-photon interference, we made no effort to further stabilise our setup to observe them.

Consider the correlations about zero delay: $a'(t)$ and its time delayed version are statistically independent so equation (7) holds but with the source correlations of $a'(t)$ instead of $a(t)$. Because the source is doubly excited the ratio of the correlations in terms of $a(t)$ is altered, which can easily be seen by applying the above rules. Now, we arrive at the measured degree of correlation
\begin{equation} \label{eq:9}
G^{(2)}_{c'd'}[0] = G^{(2)}_a [0] + \frac{1}{2} \left[ N_a^2 - \left| G^{(1)}_a [0] \right|^2 \right]
\end{equation}
choosing instead to normalise to the maximum value of this expression we arrive at the equation discussed in the main text
\begin{equation} \label{eq:10}
g^{(2)}_{MZ}[0] = \frac{2}{3}g^{(2)}_a [0] + \frac{1}{3} \left[ 1 - \left| g^{(1)}_a [0] \right|^2 \right]
\end{equation}

Next, we discuss the influence of non-equal splitting ratios at the beamsplitters to arrive at the correlations comprising each of the five observed peaks. As presented in the main text, we built our Mach-Zehnder (MZ) interferometer from fiber-coupled beam splitters and polarisation maintaining fibers. While this realization of a MZ interferometer is very stable, the commercially available fiber-coupled beam splitters have an imperfect transmittivity ($T$) to reflectivity ($R$) ratio that deviates from $50:50$. Therefore, we need to derive the amplitudes of the five peak pattern observed in the correlation measurements with the individual transmitivities and reflectivities of the first beamsplitter ($T_1, R_1$) and second beamsplitter ($T_2, R_2$) included. Both of our beamsplitters have a $T:R$ ratio of $0.56:0.44$. The transmitted component at the first beamsplitter is sent to the longer arm of the interferometer. Importantly, since this delay is realised with an additional fiber there is an additional coupling loss resulting in an effective splitting ratio of the first beamsplitter of $T_1=R_1=0.44=B$. Then, the amplitudes of the five peaks in the correlation measurements with SPCM2 as start and SPCM1 as stop are given by:
\begin{align}
& A_1 = A B^2 \, R_2^2 \\
& A_2 = A B^2 \, \left(2 R_2T_2 + 2 R_2^2 g^{(2)}_a[0]\right) \\
& A_3 = A B^2 \, \left(4 R_2T_2 g^{(2)}_a[0] + \left(R_2^2 + T_2^2 -2|g^{(1)}_a[0]|^2R_2T_2\right)\right) \\
& A_4 = A B^2 \, \left(2 R_2T_2 + 2 T_2^2 g^{(2)}_a[0]\right) \\
& A_5 = A B^2 \, T_2^2
\end{align}
where A is the absolute amplitude. These equations can now be used to fit the measured data. We determine $A B^2$ from fitting a number of peaks away from zero time delay using $g^{(2)}_a[0] = 1$ and $|g^{(1)}_a[0]|^2 = 0$, followed by fitting the five peaks around zero time delay. Note that when binning the data also an extremely weak dark-count noise was subtracted.

\begin{figure}[!t]
\includegraphics[width=\columnwidth]{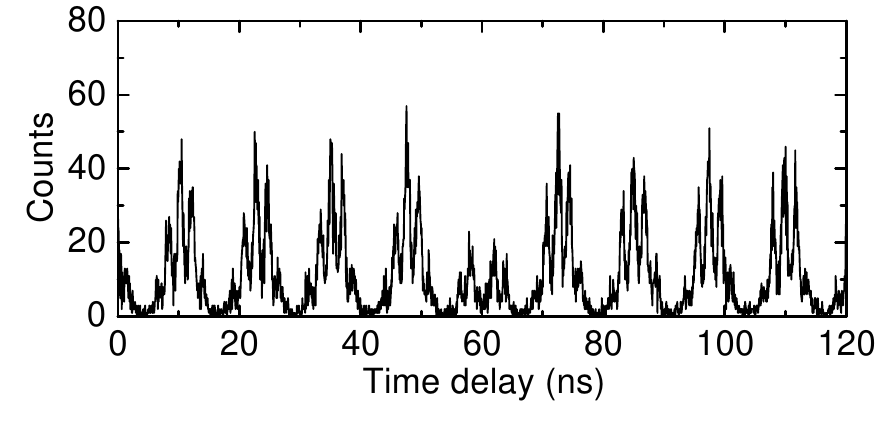}
\caption{\textbf{Two photon interference.} Raw data of the frequency filtered MZ measurements.}
\label{figure:S4}
\end{figure}

A binned version of the frequency filtered two-photon interference measurements was presented in Fig. 2d of the main text. For completeness, the raw data of this measurement is presented in Fig. \ref{figure:S4}.

\section*{\textsf{Lower limit of  $g^{(2)}[0]$}}
\begin{figure}[!t]
\includegraphics[width=\columnwidth]{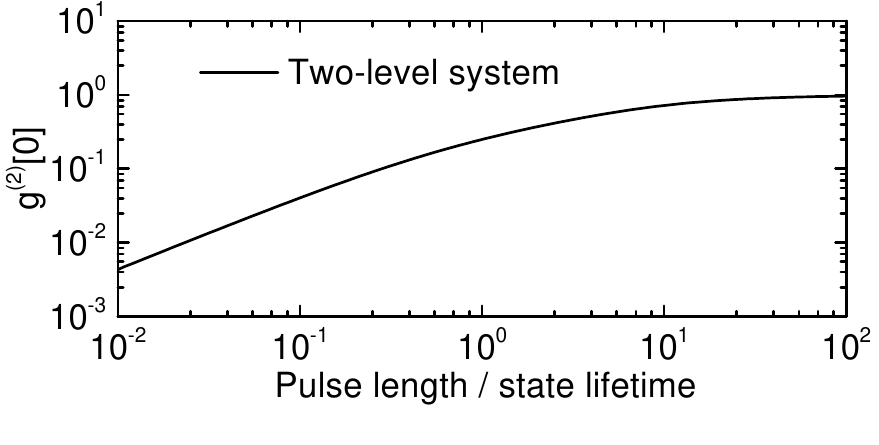}
\caption{\textbf{Influence of pulse length.} Simulation of  $g^{(2)}[0]$ obtained from a two-level system under resonant excitation with $\pi$ pulses of variable length.}
\label{figure:S5}
\end{figure}
Finally, we discuss the non-zero value of $g^{(2)}[0]$. For strongly coupled QD-cavity systems $g^{(2)}[0]$ strongly depends on the pulse length, which has to be chosen as a compromise between re-excitation during the presence of the pulse and spectral overlap of the laser with higher climbs up the JC ladder \cite{Kai2015PRX}. However, when using SHS and frequency filtering the situation simplifies and becomes identical to that of a two-level system, where the only limitation is re-excitation which is determined by the pulse length. To visualize this dependence, we present in Fig. \ref{figure:S5} a simulation of $g^{(2)}[0]$ obtained from a two-level system under resonant excitation with $\pi$ pulses as a function of the pulse length. Clearly,  $g^{(2)}[0]$ increases with pulse length and very small values are only reached for pulses that are much shorter than the state lifetime.

\end{document}